# Contact line pinning and de-pinning can modulate the rod-climbing effect

Navin Kumar Chandra,[a,⊥] Udita. U. Ghosh,[b,⊥] Aniruddha Saha,[c] Aloke Kumar [a *]

[a] *Department of Mechanical Engineering, Indian Institute of Science, Bangalore, India.*
[b] *Department of Chemical Engineering and Technology, Indian Institute of Technology (BHU), Varanasi, India.*
[c] *Department of Mechanical Engineering, Indian Institute of Technology, Kharagpur, India.*

## ABSTRACT

Our experiments on the rod-climbing effect with an oil-coated rod revealed two key differences in the rod-climbing phenomena compared to a bare rod. On one hand, an enhancement in magnitude of climbing height for any particular value of rod rotation speed and second, a decrease in threshold rod rotation speed required for the appearance of the rod-climbing effect were observed. Observed phenomena is explained by considering the contact line behavior at the rod-fluid interface. Transient evolution of meniscus at the rod-fluid interface revealed that the three-phase contact line was pinned for a bare rod and de-pinned for an oil-coated rod. We modelled the subject fluid as a Giesekus fluid to predict the climbing height. The differences in the contact line behaviour were incorporated via the contact angle at the rod-fluid interface as a boundary condition. An agreement was found between the observed and predicted climbing height establishing that contact line behaviour may modulate rod-climbing effect.

[*]Corresponding author e-mail: alokekumar@iisc.ac.in

[⊥]equal contribution

## 1. INTRODUCTION

Rod-climbing aka the Weissenberg effect was identified in the paints industry[1] and later described in scientific literature by Garner and Nissan,[1] also by Russel.[2] Weissenberg[3] postulated that such distortion in the meniscus profile is driven by the non-zero value of the first and second normal stress differences under shear flow.[4–8] Rod-climbing has been reported in a variety of fluids ranging from condensed milk,[9] hen's egg[10] to generic polymer melts.[11] Although prevalent over such a wide range of fluids that are viscoelastic in nature it is not exclusive to the class of non-Newtonian fluids.[12–14] Instances of Weissenberg effect in Newtonian fluids have been reported [15,14,16] albeit in two immiscible fluids layered horizontally. More recently, it has been shown that the Weissenberg effect can be employed for manufacturing of polymer nanofibers[17] and direct writing of polymer micropatterns.[18] In general, the parameters that modulate the height of rod-climbing can be enumerated as angular velocity of rod,[19,20,21] fluid properties[4,6,8] (surface tension, density, viscosity) and geometrical parameters[12,20] (rod radius, rod immersion length).

However, literature elaborating the effect of the rod-fluid interfacial condition on rod-climbing phenomenon is sparse. Rod-fluid interfacial condition for a particular system can be altered by changing the rod surface properties like its wettability, roughness or material composition. Zhao et al.[15] reported that rods of varied material (stainless steel, PVC, PTFE) produced comparable magnitude of climbing height in the rod-climbing of two immiscible Newtonian fluids. The authors also state based on this observation, that the rod-climbing effect is a hydrodynamic effect instead of a wetting phenomenon. However, Joseph et al.[6] reported higher magnitude of climb with aluminium rod (contact angle = $50^0$) compared to Teflon rod (contact angle = $38^0$) in their rod-climbing experiments with STP motor oil as the subject fluid. Beavers et al.[22] proposed a theory for development of rotating rod viscometer by assuming a flat meniscus at the rod-fluid interface. They achieved this neutral wetting condition experimentally by using a Scotchgard coated rod which gives a contact angle of $90^0$ with STP motor oil. Experimental data from their work has been used as a benchmark for validation of various numerical investigations[23,24,25,26] of Weissenberg effect. All these studies considered the condition of vanishing curvature (contact angle = $90^0$) at the rod-fluid interface irrespective of its validity in the physical scenario being emulated. Moreover, it is also possible that the contact angle may exhibit a range of values from a minimum to a maximum value in case of pinned contact line.[27] Thus accounting for contact line behaviour through unique contact angle is questionable. Effect of this contact line behaviour at the rod-fluid interface is an unexplored aspect of the Weissenberg effect.

Here, we investigate the role of contact line behaviour on the Weissenberg effect for a viscoelastic fluid. We replaced the bare rod with an oil-coated rod which led to the alteration of the contact line behaviour from pinned to de-pinned mode at the rod-fluid interface. The experimental observations were quantified in terms of the magnitude of the rod-climbing height, $h_c$ attained by the fluid as a function of rod rotation speed, $\omega$ ranging from 200 to 1000 RPM. We observed a significant enhancement in $h_c$ on introduction of the oil-coated rod. For instance, $h_c$ increased from ~1.3 mm to ~2.6 mm at 600 RPM. Also, we observed the existence of a threshold rod rotation speed beyond which the fluid exhibited rod-climbing. Further, we show that this threshold rod rotation speed depends on interfacial condition. We hypothesized that the observations of the present study can be explained by accounting for the contact line behaviour at the rod-fluid interface. To predict $h_c$, we have modelled the viscoelastic fluid of the present study as a Giesekus fluid. The effect of the contact line behaviour on $h_c$ has been incorporated as a boundary condition depending on the mode (pinned/de-pinned) of contact line dynamics. To the best of our knowledge, the present study is the first of its kind where we revisit the classical rod-climbing and outline the effect of contact line behaviour.

## 2. Materials and Methods

**2.1 Experimental parameters** - Aqueous polymeric solution of polyethylene oxide (PEO) has been employed as a model fluid in the present study. This solution of PEO powder (viscosity average molecular weight ~ $5 \times 10^6$ g/mol, *Sigma Aldrich*) was prepared by stirring (500-700 RPM for 72 hours at ambient temperature, 25˚C) in deionized water to obtain a clear i.e., an optically transparent solution of concentration 1% (w/w). Rod-climbing occurred at 1% (w/w) when the angular speed of rotation, $\omega$ provided by the overhead stirrer was ramped externally from 200 to 1000 RPM in increments of 200 RPM. The main objective was to explore the response of a reference viscoelastic fluid in presence of oil-dipped rod. Therefore, the reference fluid concentration was fixed at 1% (w/w). Other parameters like rod diameter, $d$ ~ 1 cm, fluid level (~ 4 cm) in the tank, rod immersion depth (~ 3.5 cm) and the operating temperature (25 ºC) were also kept fixed.

**2.2 Characterisation of reference fluid** - The rheology of the polymeric solution and the viscosity of the silicone oil were characterized using a cone and plate geometry of a rheometer (*Model: MCR302, Anton Paar*). Surface tension of the polymeric solution and silicone oil was measured using a Goniometer (*Model: 250G1, Ramèhart, Germany*). All the characterizations were performed at room temperature (25 ºC).

**2.3 Experimental setup** - The present experimental setup comprises of a cuboid acrylic tank of dimensions ($H = W = L = 10$ cm) as shown in Figure 1(a), placed symmetrically with respect to an overhead stirrer.

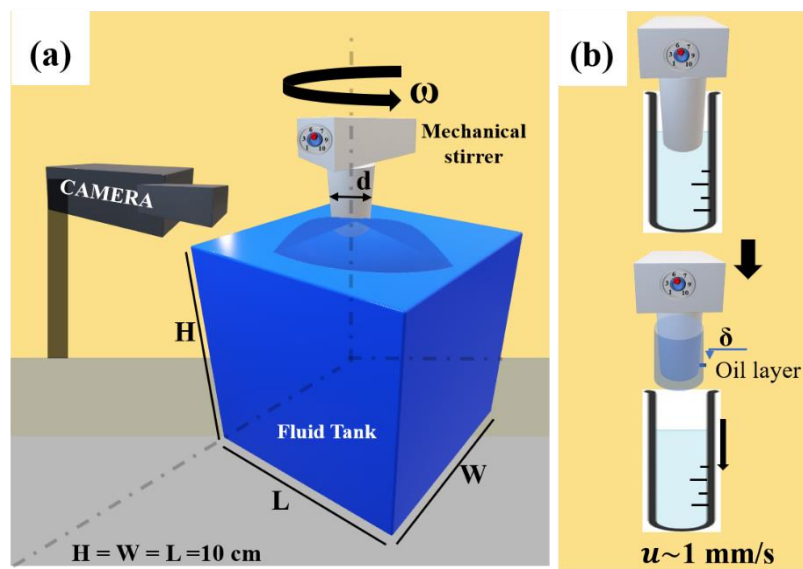

**Figure 1 (a)** Schematic of the experimental setup, a rotating rod with diameter $d = 1\ cm$ and rotation speed $\omega$ is partially inserted in the polymeric solution. **(b)** Process of introducing an oil layer of thickness δ, at the rod-fluid interface with oil reservoir withdrawal velocity, $u \sim 1mm/s$.

A DSLR camera (*Model: D850, Nikon*) with Navitar zoom lens was carefully aligned with the stirrer axis to track the free surface profile. To clearly visualize the free surface profile near the rod-fluid interface, the polymeric fluid was dyed with Phenol red (*Sigma Aldrich*) and thereafter it was poured into the fluid tank. Prior to start of the experiment, a reference line was marked on the rod (figure 2). The stirrer is activated once a stable static meniscus of the PEO solution was attained. Stirrer was kept rotating at a constant RPM for ~30 seconds and then brought to a halt. This duration was sufficient for the polymeric fluid to climb up along the rod and attain a steady profile. After the rod rotation was brought to a halt, sufficient time (~ 5 minutes) was allowed for the meniscus to recede completely and attain the final equilibrium position. The whole sequence from static to rotating to final recession of the meniscus was captured using the DSLR camera corresponding to an angular speed of rotation, $\omega$. We refer to this whole cycle as one experimental dataset and a minimum of three datasets were used for error analysis and averaging. Using the same procedure, experiments were performed with bare rod and oil-coated rod. Oil layer coating on the rod surface was achieved by dipping the rod into a silicone oil reservoir and then withdrawing it in a controlled manner (Figure 1b). The velocity of rod withdrawal from oil reservoir, $u \sim 1 mm/s$ was kept constant. Surface of the rod was cleaned with isopropyl alcohol prior to its immersion into the oil while acquiring every new dataset. The silicone oil (density, $\rho_{oil} = 0.968\ g/cm^3$) was characterized by measuring its surface tension ($\gamma_{oil} \sim 21\ mN/m$) and dynamic viscosity ($\mu_{oil} \sim 355\ mPa.s$). The corresponding capillary number, $Ca$ $\left(Ca = \frac{\mu_{oil} u}{\gamma_{oil}}\right)$ was found to be 0.01. Since, $Ca \ll 1$ an approximate thickness, $\delta$ (mm) of the oil layer present on the oil-dipped rod can be estimated by the Landau-Levich[28] equation, $\delta \sim 0.94 \frac{(\mu_{oil} u)^{2/3}}{(\rho_{oil} g)^{1/2} (\gamma_{oil})^{1/6}}$ and was estimated to be ~ 0.09 mm for the present study.

**2.4 Image processing** – For better visualization, figures like Figure (2), are presented here with an artificially created background. All image processing however was performed on RAW and unaltered images. Camera RAW image files were analysed using ImageJ (*NIH, version: 1.53e*) software for experimental evaluation of different quantities like climbing height and contact angle.

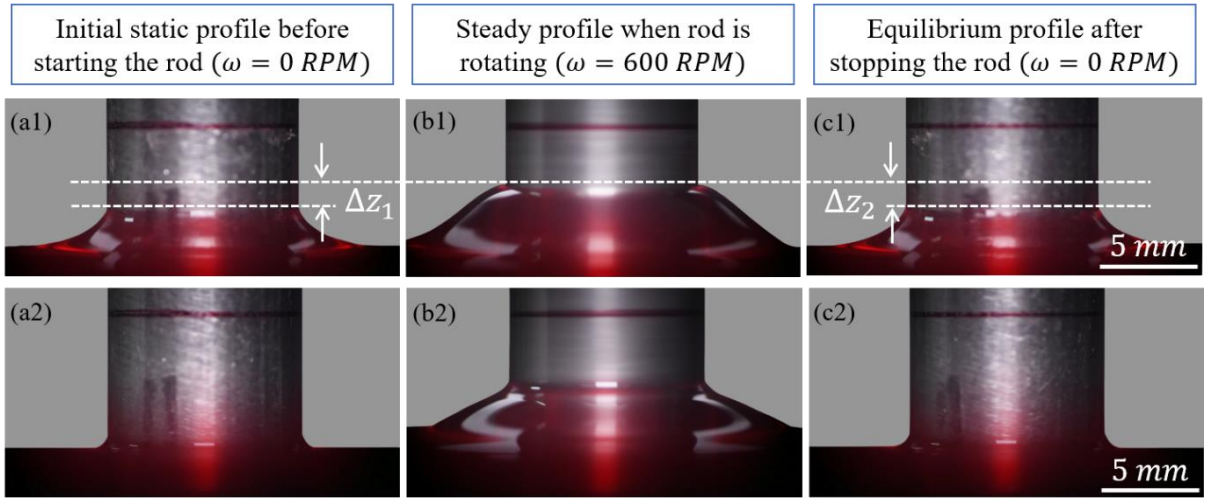

**Figure 2** Experimental images showing the free surface profile with bare rod and oil-coated presented in first and second row, respectively. **(a1, a2)** initial static profile before starting the rod rotation, **(b1, b2)** steady state profile corresponding to rod rotating at 600RPM **(c1, c2)** receded equilibrium profile after the rod is stopped.

The climbing height was quantified by two differences (Figure 2). The difference in steady state vertical position of three phase contact line when the rod is in rotating condition taken with respect to: (i) the initial static meniscus before rotating the rod, $\Delta z_1$ and (ii) the final receded equilibrium meniscus after stopping the rod, $\Delta z_2$. It was found that the two differences were almost equal for a particular dataset and therefore, average of both the differences is hereafter reported in this study as climbing height, $h_c$. It was further non-dimensionalized, $h_c^* = h_c/d$ by the rod diameter, $d$.

## 3. Theoretical Formulation

The experimental setup in the present study is analogous to the classical Taylor-Couette device comprising of two concentric cylinders with inner and outer radii, $R_i = 5 \times 10^{-3}\ m$ and $R_o = 5 \times 10^{-2}\ m$ respectively such that only the inner cylinder rotates at a speed, $\omega$ while the outer cylinder is stationary. The flow of the polymeric solution in the annular space is assumed to be incompressible and purely azimuthal for present experimental range of angular velocity of rod. We employ cylindrical coordinate system with $r, \theta, z$ as the coordinate axes. $r$ is the radial direction along the rod radius, $\theta$ is azimuthal direction and $z$ is the vertical direction along the axis of the rod. $z = h(r, \omega)$ presents the steady free surface profile of the polymeric fluid in the annular gap ($R_i \leq r \leq R_o$) corresponding to the rod rotation speed, $\omega$. Static profile is given by $z = h(r, \omega = 0)$ when the rod is in static condition. We have defined climbing height as the vertical displacement of the three-phase contact line considering static profile as the datum therefore, climbing height $h_c$ can be calculated using equation (1).

$$h_c = h(r = R_i, \omega) - h(r = R_i, \omega = 0) \qquad (1)$$

At steady state, the general governing equation for the free surface profile[29] of a fluid subjected to Taylor-Couette flow with only inner cylinder rotation is given by equation (2).

$$(-\rho g)\frac{dh}{dr} + \frac{d}{dr}\left[\frac{\sigma}{r}\frac{d}{dr}\left(r\frac{dh}{dr}\right)\right] = \frac{N_1}{r} - \frac{dN_2}{dr} - \rho\frac{u_\theta^2}{r} \qquad (2)$$

Where $h$ is the free surface profile, $u_\theta$ is the magnitude of fluid velocity in the $\theta$ direction, $\rho$ is the fluid density and $\sigma$ is the surface tension of the fluid. First and second normal stress differences are given by $N_1 = \tau_{\theta\theta} - \tau_{rr}$ and $N_2 = \tau_{rr} - \tau_{zz}$ where $\tau_{\theta\theta}, \tau_{rr}, \tau_{zz}$ are the normal stresses in $\theta, r$ and $z$ directions, respectively. The unknowns in equation (2) to obtain the free surface profile are the velocity distribution and the normal stress differences across the annular gap. These were estimated by modelling the polymeric fluid having rheological properties governed by the Giesekus constitutive relation. Figure (3a) shows the viscosity variation of the polymeric fluid with strain rate. Figure (3b) shows the variation of storage and loss modulus with varying angular frequency in small amplitude oscillatory shear (SAOS) test. These measurements revealed the shear thinning nature and viscoelastic behaviour of the polymeric fluid, both of which can be captured by the Giesekus model.

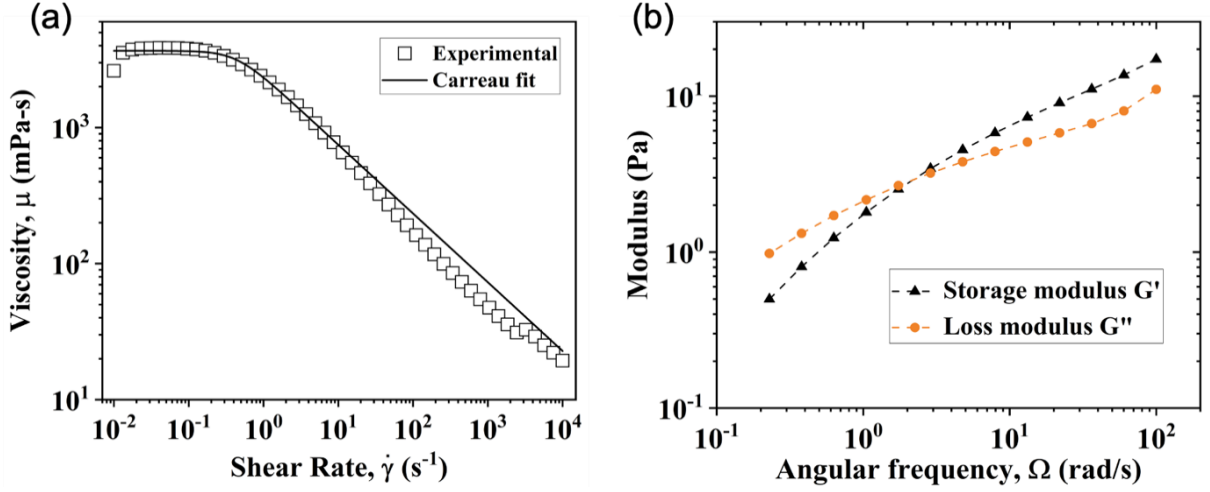

**Figure 3 (a)** Variation of viscosity $\mu$ with shear rate $\dot{\gamma}$ for the present polymeric fluid. Solid line is Carreau fit to the experimental data (fitting parameters available in supplementary information, Note S2). **(b)** Variation of storage modulus, $G'$ and loss modulus, $G''$ with angular frequency $\Omega$ in SAOS test.

Constitutive equation for a Giesekus fluid with negligible solvent viscosity[30] is given by equation (3).

$$\boldsymbol{T} + \lambda\left(\frac{\partial \boldsymbol{T}}{\partial t} + \boldsymbol{u}.\boldsymbol{\nabla T}\right) - \lambda[\boldsymbol{T}.\boldsymbol{\nabla u} + (\boldsymbol{\nabla u})^T.\boldsymbol{T}] + \frac{\lambda\alpha}{\mu_0}\boldsymbol{T}.\boldsymbol{T} = \mu_0[\boldsymbol{\nabla u} + (\boldsymbol{\nabla u})^T] \quad (3)$$

Here $\boldsymbol{u}$ is the velocity vector and $\boldsymbol{T}$ is the extra-stress tensor. There are three parameters in the Giesekus model: $\lambda$ is the stress relaxation time, $\alpha$ is the dimensionless mobility parameter and $\mu_0$ is the polymer contribution to the zero-shear viscosity of polymeric solution. Dapra and Scarpi[30] have shown that purely azimuthal Taylor-Couette flow of a Giesekus fluid can be solved analytically to get the normal stress differences and the velocity profile. Analytical expression for $u_\theta$, $N_1$ and $N_2$ were directly taken from this literature (Refer to the supplementary information, Note S2 for detailed expression). Equation (2) was then integrated from some arbitrary radial location, $r$ to $R_o$ within the annular gap resulting in a second order differential equation in $h$,

$$\rho g h - \rho g h(R_o) - \sigma\left[\frac{d^2h}{dr^2} + \frac{1}{r}\frac{dh}{dr}\right] + \sigma\frac{d^2h}{dr^2}\bigg|_{R_o} + \frac{\sigma}{R_o}\frac{dh}{dr}\bigg|_{R_o} = \int_r^{R_o}\left(\frac{N_1}{r} - \frac{\rho u_\theta^2}{r} - \frac{dN_2}{dr}\right)dr \quad (4)$$

Equation (4) was further simplified by imposing the condition $\int_{R_i}^{R_o} rh\, dr = 0$ which ensures that the bulk fluid mass remains conserved. This equation was numerically solved using 4th order Runge-Kutta scheme with the following appropriate boundary conditions,

$$h'(r = R_i) = -\cot(\theta_i) \quad (5a); \quad h'(r = R_0) = \cot(\theta_0) \quad (5b)$$

where $\theta_i$ and $\theta_0$ are the contact angles at inner and outer boundaries, respectively. Equation (4), here on referred to as the governing equation can be solved subjected to boundary condition given in equation (5), to obtain the free surface profile. Climbing height can be then calculated using equation (1).

While applying boundary conditions, $\theta_0$ can be taken as 90° to ensure flat surface far away from the rod. Since most of the hydrodynamic effects and hence, surface deformation appears near the rod, therefore $\theta_i$ becomes an important parameter in determining the climbing height. In case of pinned contact line, $\theta_i$ is not unique and it can take values within a given range. If $\theta_i$ becomes a variable then equation (5a) cannot be used as a boundary condition, hence solution of the governing equations to get the free surface profile is not straight forward. The experimental evidence of variability of $\theta_i$ and proper solution procedure is discussed in the next section.

## 4. Results and Discussions

Classical rod-climbing reported widely in polymeric systems has also been observed here with bare rod as well as oil-coated rod. However, introduction of oily rod led us to record two key observations. First, an increase climbing height for a particular rod rotation speed. Second, for rod rotating at 200 RPM, no climbing was observed with bare rod whereas oil-coated rod shows significant amount of climbing. This suggest that there exists an interfacial condition dependent threshold rotation speed, $\omega_{th}$ below which rod-climbing will not occur. Introduction of oily rod decreases $\omega_{th}$. These observations can be explained by accounting for the contact line behaviour at the rod-fluid interface. It was observed that contact line is in pinned mode with bare rod and in de-pinned mode with oil-coated rod. Figure (4) shows the transient behaviour of contact line when the rod was brought to a halt after rotating at 800 RPM. The hydrodynamic forces started vanishing when the rod was brought to halt, as a result the climbed-up fluid started to recede downward under the action of gravitational force. It can be observed that as the bare rod was brought to a stop (Figure 4a), first response of fluid appears as the change in the contact angle while the contact line remains pinned at a fixed location. It continues to remain pinned till the contact angle reaches a minimum value and then the contact line starts to recede. Whereas, in case of oil-coated rod (Figure 4b), the contact line starts to recede immediately and a rapid increment in vertical displacement of the contact line, Δz was observed after stopping the rod (See the supplementary video, S1 for details).

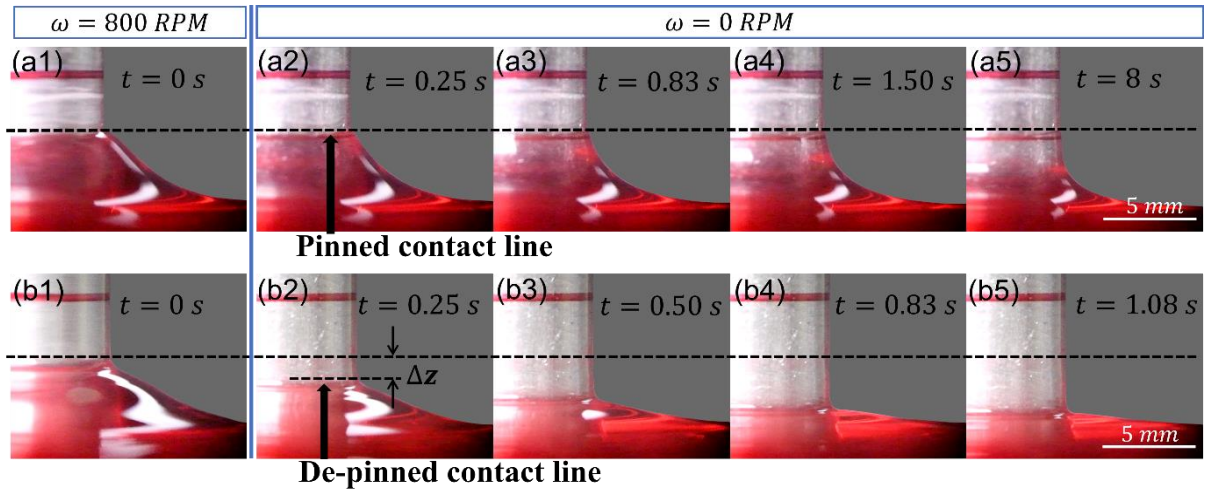

**Figure 4** Experimental images showing temporal evolution of meniscus for the **(a)** Pinned and **(b)** de-pinned contact line with bare rod and oil-dipped rod, respectively. Only right half of the axisymmetric profile is shown. Here, panel 1 corresponds to rod in rotating condition (ω = 800 RPM) and panels 2,3,4,5 depict the meniscus evolution once the rod rotation is halted. t = 0 second corresponds to the time instant just before stopping the rod. Vertical blue line demarcates the rotating and the static condition of the rod.

It takes ~10 seconds for the contact line to recede over an oil-coated rod and attain the equilibrium position. Whereas the contact line over bare rod remains pinned for ~12 seconds and takes comparatively longer time (~60 seconds) to attain the final receded equilibrium position. These observations led us to model the contact line of polymeric fluid with bare rod in pinned mode and with oil-coated rod in de-pinned mode. To incorporate the effect of contact line behaviour in determining the climbing height, we propose a solution procedure based on the mode of contact line dynamics as follows:

***Pinned contact line***: In case of pinned contact line, the contact angle, $\theta_i$ is not unique, rather it varies within a range of values. This range can be defined by the minimum, $\theta_{min}$ and the maximum, $\theta_{max}$ value of contact angle at the rod fluid interface. $\theta_{min}$ was observed when the rod was in static condition ($\omega = 0$) and $\theta_{max}$ was observed when the rod was rotating above the threshold rotation speed ($\omega \geq \omega_{th}$). The origin of threshold rotation speed can be traced to the existence of inherent surface (rod) hysteresis that the moving wetting line of the polymeric fluid has to overcome to attain a climb. This non-zero climbing height with a pinned contact line for $\omega \geq \omega_{th}$ can be calculated from equation (6).

$$h_c^p = h(r = R_i, \omega \geq \omega_{th}) - h(r = R_i, \omega = 0) \qquad (6)$$

Here, superscript $p$ stands for pinned mode of contact line. First term on the RHS of equation (6) needs to be obtained using equation (4) and boundary condition (5a) such that $\theta_i = \theta_{max}$. Second term can be evaluated in the same manner but keeping $\theta_i = \theta_{min}$. At $\omega = \omega_{th}$ the hydrodynamic forces are just sufficiently balanced by the pinning force. Therefore equation (6) can also be used to estimate the threshold rotation speed, $\omega_{th}$ by using the condition that $h_c^p = 0$ at $\omega = \omega_{th}$. At $\omega < \omega_{th}$, the climbing height is always zero because at these rotation speeds the hydrodynamic forces are not sufficient to overcome the pinning force and to cause the upward motion of contact line. At this condition, the hydrodynamic forces will be balanced only by the change in curvature of the free surface profile. For example, at $\omega = 200\ RPM$ with bare rod, the climbing height was zero, but the free surface curvature was altered due to rod rotation (Figure 5). For the rod rotation range of $0 < \omega < \omega_{th}$, $\theta_i$ becomes a variable and its value lies between $\theta_{min}$ and $\theta_{max}$. In this region of rod rotation, equation (5a) is no longer a boundary condition rather it becomes an equation to evaluate $\theta_i$ along with simultaneously satisfying the condition $h_c^p = 0$.

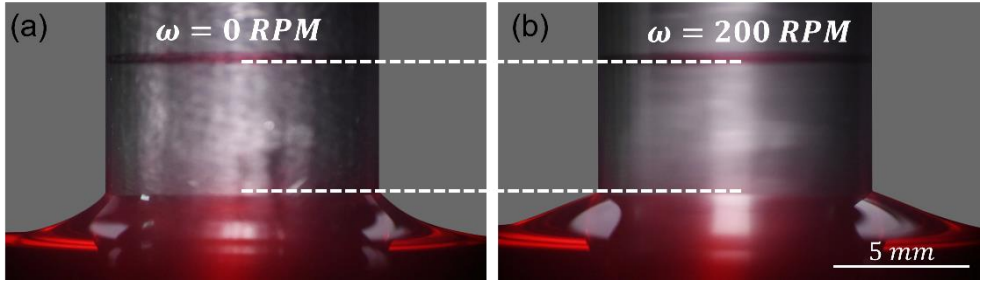

**Figure 5** Experimental images showing steady free surface with bare rod **(a)** in static condition, **(b)** rotating at 200 RPM. Upper white dashed line indicates the vertical position of reference line marked on the rod. Lower white dashed line indicates the vertical position of contact line at the rod-fluid interface.

***De-pinned contact line***: In case of de-pinned contact line, the contact angle $\theta_i$ have a unique value $\theta_{dp}$, which is independent of rod rotation speed. Since, pinning force is absent, hydrodynamic forces due to any non-zero rod rotation will result in non-zero climbing height. Hence, $\omega_{th}$ for this case will be zero. Calculation of climbing height with de-pinned contact line, $h_c^{dp}$ is straight forward and it can be obtained from equation (1) by letting $h_c = h_c^{dp}$. Unlike pinned mode, there is no complexity in solving for $h_c^{dp}$ because the $\theta_i$ remains constant for all $\omega$.

Thus, the de-pinned mode on oil-coated rods and the pinned mode on bare rod are analogous to the constant contact angle (CCA) and constant contact radius (CCR) modes resp. observed in droplet evaporation. Briefly, in CCA mode, the contact radius changes as evaporation proceed but the contact angle remains constant. While in CCR mode, the interfacial contact area remains constant, and the contact angle becomes a varying parameter.

The free surface profile, $h(r, \omega)$ and climbing height, $h_c$ were obtained theoretically and shown in figure (6) and (7) respectively. Figure (6a) presents the theoretically obtained free surface profile with bare rod (modelled as pinned contact line) at various rotational speed. It can be observed that for $0 \leq \omega \leq \omega_{th}$, contact angle $\theta_i$ varies with rod rotation speed whereas the contact line remains pinned at the rod surface. This explains the zero height of climbing for $0 \leq \omega \leq \omega_{th}$. For $\omega > \omega_{th}$, the contact angle remains constant, but the climbing height increases with increasing $\omega$ of the rod. Figure (6b) and (6c) shows the theoretically obtained free surface profile overlapped on top of the experimental images.

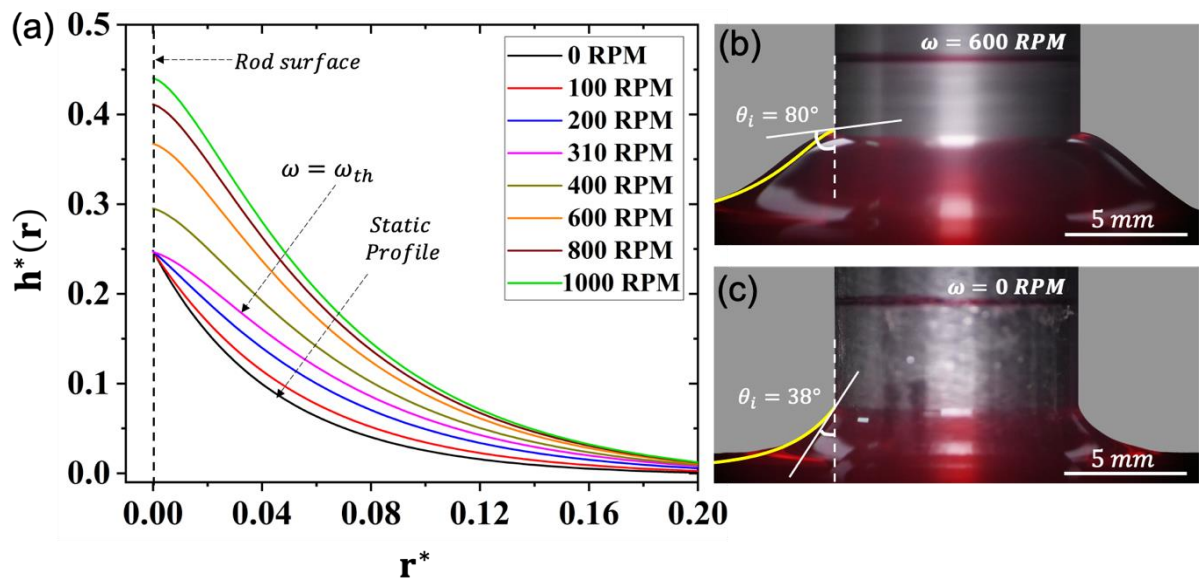

**Figure 6 (a)** Theoretically obtained free surface profile for different rotational speed (ω) of the rod. $h^* = h/d$ is the non-dimensionalized height profile and $r^* = (r - R_i)/(R_o - R_i)$ is the non-dimensionalized radial position. Theoretical profile (yellow curve) is superimposed on the experimental images corresponding to the bare rod **(b)** rotating at 600 RPM and **(c)** in static condition, respectively.

For the purpose of theoretical calculation, different parameters in the governing equation needs to be evaluated. $\theta_{min}$ and $\theta_{max}$ for bare rod were obtained from image analysis of experimental images and their average values were found to be ~38° and ~80° respectively. Contact angle with oil-coated rod (de-pinned mode), $\theta_{dp}$ has been taken as 45°. It is difficult to measure $\theta_{dp}$ due to presence of oil layer. However, exact value of $\theta_{dp}$ do not have significant effect on climbing height because it is a constant for both the conditions, i.e., rotating and static, and hence their difference remains unchanged. Surface tension of the polymeric fluid, $\sigma$ was measured using the Goniometer and found to be 56 mN/m. Polymer contribution to the zero-shear viscosity, $\mu_0 = 3.67\ Pa\text{-}s$, was obtained from the Carreau fit (Refer supplementary information for fitting parameters, Note S2) to the experimental data of viscosity variation with shear rate as shown in figure (3a). It was found that $\lambda = 0.04\ s$ and $\alpha = 0.216$ provides best match between theory and experimental data of the present study. It should be noted that $\lambda$ and $\alpha$ are material properties of polymeric fluid and therefore were kept unaltered in all the theoretical calculations. Figure (7) presents the comparison of experimental and theoretical climbing height. Theoretical values are in good agreement with experimental results for bare rod at lower RPM ($\omega \leq 800\ RPM$). Mismatch at 1000 RPM could be due to violation of purely azimuthal flow at higher rotational speed. In case of oil-coated rod, theory is over-predicting the climbing height. Possible reason for this mismatch could be due to modelling of perfectly de-pinned contact line with oily rod. Whereas some amount of pinning may be present in actual condition.

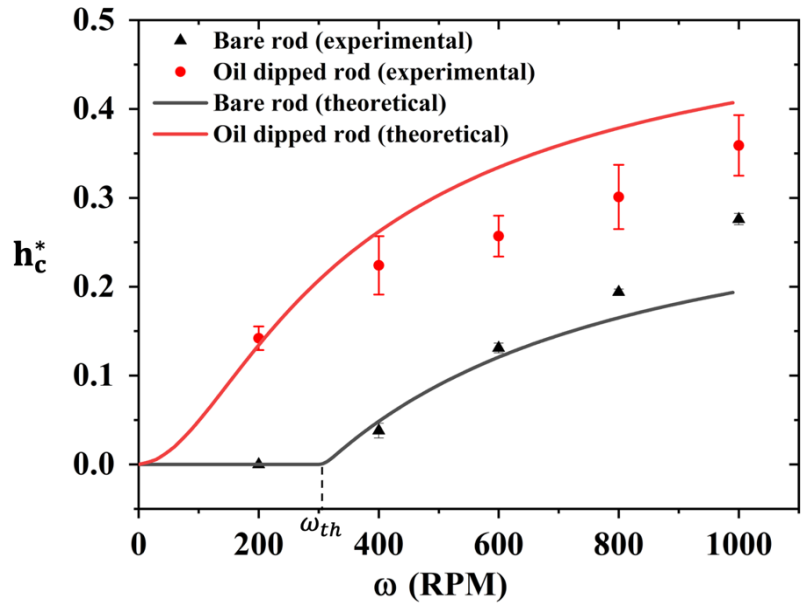

**Figure 7** Comparison of the predicted and experimental value of the non-dimensionalized climbing height, $h_c^* = h_c/d$ for the polymeric fluid meniscus with bare rod and in presence of oil-coated rod.

Despite some mismatch between theoretical and experimental values, proposed theory can explain the two key experimental observations of the present study. It is clear from figure (7) that the climbing height will be always higher in case of de-pinned mode compared to pinned mode of contact line. This explains the experimental observation of enhancement in the climbing height on introduction of oil-coated rod. The proposed model also predicts the threshold rod rotation speed, $\omega_{th}$ required for the occurrence of rod-climbing effect. As per the mechanism proposed in the present study, $\omega_{th}$ stems from the pinning of the contact line at the rod-fluid interface. For bare rod in the present study, $\omega_{th}$ from equation (6) was found to be 310 RPM. This explains the experimental observation of absence of climb at 200 RPM in case of bare rod. But when the bare rod was replaced by the oil-coated rod, the contact line de-pins leading to measurable climb even at 200 RPM. In other words, oil layer assists rod-climbing by de-pinning the contact line. To further understand the effect of contact line pinning on climbing height, we used contact angle hysteresis, $CAH$ as the parameter to quantify the amount of pinning such that $CAH = \theta_{max} - \theta_{min}$. Figure (8) presents the variation of theoretically obtained climbing height with varying rod rotation speed and $CAH$.

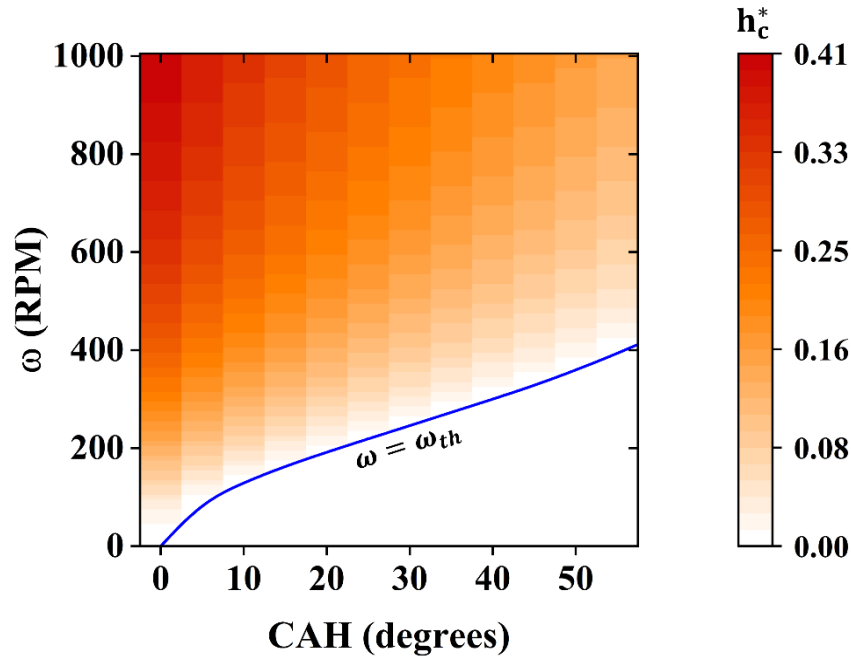

**Figure 8** Colourmap showing the variation of non-dimensionalized climbing height, $h_c^*$ with rod rotation speed, $\omega$ and contact angle hysteresis, $CAH$. Here, $CAH$ corresponds to $\theta_{min} = 38°$ as in the case of bare rod of the present study. The blue line denotes the threshold rotation speed, $\omega = \omega_{th}$ as a function of $CAH$.

Two distinct regions can be seen in Figure 8 viz., coloured and white region. Colours in coloured region are mapped with the non-dimensionalized climbing height, $h_c^* = h_c/d$. White area represents the region of no climbing. The two regions are demarcated by the line, $\omega = \omega_{th}$ as a function of $CAH$. Below this line rotational speed of the rod is not sufficient to de-pin the contact line and hence the climbing height is zero. It can be observed that $\omega_{th}$ increases with amount of pinning. To the best of our knowledge, none of the existing literatures have reported the existence of threshold rotation speed for rod-climbing of viscoelastic fluids. Although existence of $\omega_{th}$ has been reported for the rod-climbing effect of two immiscible Newtonian liquids[14], the governing physics is completely different from the rod-climbing of viscoelastic fluids. Existence of $\omega_{th}$ for the former is related with the onset of Taylor-Couette instability in Newtonian fluid. Whereas, proposed existence of $\omega_{th}$ in viscoelastic fluids is related with the contact line pinning at the rod-fluid interface. The present work dealt with the significance of contact line behaviour, applicable only to steady state meniscus profile achieved during the rod-climbing of viscoelastic fluids. Further research is required to incorporate the effect of contact line behaviour on the transient evolution of meniscus profile in the rod-climbing effect.

## 5. CONCLUSIONS

Here we studied the classical rod-climbing effect of viscoelastic fluid by replacing the bare rod with an oil-dipped rod. We experimentally showed that there are two aspects that are affected by introducing oil-dipped rods. First, the magnitude of climbing height increased and second, the threshold rod rotation speed required for the appearance of rod-climbing effect is reduced. Here, we reported the existence of an interfacial condition-dependent threshold rod rotation speed for rod-climbing of viscoelastic fluid. These observations have been explained by accounting for the contact line behavior at the rod fluid interface. Time evolution of free surface profile revealed that the three phase contact line exhibits pinned mode with bare rod and de-pinned mode with oil-dipped rod. These two different modes of contact line behavior were the underlying cause for the observed differences in rod-climbing effect with oil-dipped rod. We modelled the viscoelastic fluid as a Giesekus fluid to predict the climbing height. Suitable contact angles in terms of boundary condition have been proposed which accounts for the change in contact line behavior on introduction of oil-coated rod. Although rod-climbing effect is a classical phenomenon, this effect has potential applications in various fields. Recently it has been employed to design an experiment which mimics the dynamics of accretion disk around black holes.[31] Present study outlines the importance of contact line behavior in the rod-climbing effect of viscoelastic fluids.

## ACKNOWLEDGEMENTS

All the authors would like to acknowledge Dr. Abhijit Chandra Roy, DST Inspire Fellow, Department of Physics, Indian Institute of Science Bangalore, for building a customized dip coater used in the experiments. UUG acknowledges support from the C.V. Raman Postdoc Fellowship. NKC acknowledges support from the Prime Ministers Research Fellowship (PMRF). AK acknowledges support from DST-SERB grant no. EMR/2017/003025.**CONFLICTS OF INTEREST**

There are no conflicts to declare.